 \documentclass[preprint2]{aastex}

\shorttitle{Hot debris dust around HD~106797}
\shortauthors{Fujwiara et al.}

\begin{document}

\title{Hot debris dust around HD~106797}


\author{
Hideaki~Fujiwara\altaffilmark{1}, 
Takuya~Yamashita\altaffilmark{2}, 
Daisuke~Ishihara\altaffilmark{3}, 
Takashi~Onaka\altaffilmark{1}, 
Hirokazu~Kataza\altaffilmark{3}, 
Takafumi~Ootsubo\altaffilmark{3}, 
Misato~Fukagawa\altaffilmark{4}, 
Jonathan~P.~Marshall\altaffilmark{5}, 
Hiroshi~Murakami\altaffilmark{3}, 
Takao~Nakagawa\altaffilmark{3}, 
Takanori~Hirao\altaffilmark{3}, 
Keigo~Enya\altaffilmark{3},
and Glenn~J.~White\altaffilmark{5,6}
}


\altaffiltext{1}{Department of Astronomy, School of Science, University
of Tokyo, Bunkyo-ku, Tokyo 113-0033, Japan; fujiwara@astron.s.u-tokyo.ac.jp}
\altaffiltext{2}{National Astronomical Observatory of Japan, 2-21-1 Osawa, Mitaka, 
Tokyo 181-0015, Japan}
\altaffiltext{3}{Institute of
Space and Astronautical Science, Japan Aerospace Exploration Agency, 3-1-1 Yoshinodai, 
Sagamihara, Kanagawa 229-8510, Japan}
\altaffiltext{4}{Graduate School of Science, Osaka University, 1-1 Machikaneyama, 
Toyonaka 560-0043, Osaka, Japan}
\altaffiltext{5}{Department of Physics and Astronomy, The Open University, Walton Hall, Milton Keynes, MK7 6AA, England}
\altaffiltext{6}{Space Science \& Technology Department, The Rutherford Appleton Laboratory, Chilton, Didcot, Oxfordshire OX11 0QX, England}


\begin{abstract}
Photometry of the A0~V main-sequence star HD 106797 with {\it AKARI} and Gemini/T-ReCS 
is used to detect excess emission over the expected stellar photospheric emission 
between 10 and $20~\micron$, which is best attributed to hot circumstellar debris 
dust surrounding the star.
The temperature of the debris dust is derived as $T_{\rm d} \sim 190$~K 
by assuming that the excess emission is approximated by a single temperature blackbody. 
The derived temperature suggests that the inner radius of the debris disk is $\sim 14$~AU.
% Rin=10AU
The fractional luminosity of the debris disk is 1000 times brighter 
than that of our own zodiacal cloud. 
The existence of such a large amount of hot dust around HD~106797 cannot be 
accounted for by a simple model of the steady state evolution 
of a debris disk due to collisions, and it is likely that transient events
play a significant role. 
Our data also show a narrow spectral feature between 11 and $12~\micron$
attributable to crystalline silicates, 
suggesting that dust heating has occurred during the formation and evolution 
of the debris disk of HD~106797.
\end{abstract}




\section{Introduction}

{\it Infrared Astronomical Satellite} ({\it IRAS}) detected a number of main-sequence stars 
that show infrared excesses above their expected photospheric emission 
\citep[e.g.][]{aumann84,rhee07}.
These infrared excesses are thought to originate from second generation dust grains 
formed as a consequence of the collision of planetesimals, 
or the destruction of cometary objects \citep[e.g.][]{backman93,lecavelier96}.

Most of the known debris disks only show excesses at wavelengths longer than $25~\micron$. 
The excess comes from the thermal emission of dust grains with low temperatures 
($T_{\rm dust} \sim 100$~K) that exist far from the central star. 
To date, little is known about the properties of the debris disk material located 
close to the star, which has a more direct link to the formation of terrestrial planets 
than the low temperature debris \citep{meyer08}.
The recent availability of high-sensitivity surveys at $10-20~\micron$ allows 
the properties of this inner debris disk material to be measured. 
In this Letter, we report observations obtained with the mid-infrared (MIR) 
all-sky survey made by the {\it AKARI} satellite \citep{murakami07}. 

{\it AKARI} is a Japanese infrared satellite mostly dedicated 
to an infrared all-sky survey, which was launched in February 2006.
The MIR all-sky survey was performed using 9 and $18~\micron$ broad band filters 
with the InfraRed Camera (IRC) onboard {\it AKARI} until August 2008 \citep{ishihara06}. 
Since the peak of the thermal emission from hot dust grains with $T_{\rm dust} \gtrsim 200$~K
comes to around $10-20~\micron$, the {\it AKARI}/IRC all-sky survey data are a powerful tool 
to search for hot debris disks which should be connected with the formation process 
of terrestrial planets. 

Here we report a discovery of significant $18~\micron$ excess towards 
the A0~V main-sequence star HD~106797 from the {\it AKARI}/IRC all-sky survey data.
The distance to the star from the Sun is measured as $d=96 \pm 3$~pc based on {\it Hipparcos}
observations \citep{vanleeuwen07}. In addition, we discover significant excess emission 
at 11.7 and $12.4~\micron$ by narrow band photometric observations with the Gemini/T-ReCS. 

In this Letter, we show the spectral energy distribution (SED) in the MIR region 
of HD~106797 and discuss the spatial distribution and mineralogical properties 
of the hot debris dust around HD~106797.

\section{Observations and Data Reduction}

\subsection{{\it AKARI}/IRC all-sky survey}

The S9W ($9~\micron$) and L18W ($18~\micron$) images of HD~106797 were taken as part 
of the All-Sky Survey observations. The star was observed on 2006 August 6, 
2007 February 1, 2, August 6 and 7. 
In the All-Sky Survey, the IRC was operated in the scan mode with the 
scan speed of $215\arcsec$ sec$^{-1}$ and the data sampling time of 0.044 sec, 
which provided a spatial resolution of $\sim 10\arcsec$ along the scan direction. 
The spatial resolution along the cross-scan direction was $\sim 10\arcsec$ \citep{ishihara06}. 
The All-Sky Survey data were reduced using the standard {\it AKARI} pipeline software 
version 061210. 
The data from the three periods were median combined. 
The 5-$\sigma$ sensitivity for a point source per scan is estimated to be 50~mJy 
in the S9W band and 120~mJy in the L18W band, and the absolute uncertainty in flux density 
is 7~\% for the S9W band and 15~\% for the L18W band at present. 
The spatial resolution is improved to $5\arcsec$ in the both bands 
by combining the dithered data of multiple observations in the pipeline. 
The absolute position accuracy is estimated to be $5\arcsec$ \citep{ishihara06}.
The fluxes at the three periods agree with each other within the uncertainty, 
indicating no significant variations in the flux. 
HD~106797 was selected as a probable candidate of debris disk with $18~\micron$ excess 
from a preliminary search for debris disks based on the {\it AKARI}/IRC All-Sky Survey data.

\subsection{Ground-based follow-up observations with Gemini/T-ReCS}

HD~106797 was observed with the T-ReCS \citep{telesco98}, mounted on the 8 m 
Gemini South Telescope on 2007 June 10 and 12. 
Imaging observations in the 8.8 micron ($\Delta\lambda =0.8~\micron$), 
$9.7~\micron$ ($\Delta\lambda =0.9~\micron$), 
$10.4~\micron$ ($\Delta\lambda =1.0~\micron$), $11.7~\micron$ ($\Delta\lambda =1.1~\micron$), 
$12.3~\micron$ ($\Delta\lambda =1.2~\micron$), and $18.3~\micron$ ($\Delta\lambda =1.5~\micron$) 
bands were carried out. The pixel scale was $0\farcs09$ pixel$^{-1}$. 
To cancel out the background radiation, the secondary mirror chopping and 
the telescope nodding method were used. 
We used a standard star (HD~110458) from \citet{cohen99} as a flux calibrator 
and reference point-spread functions (PSFs) were derived by observations.
We observed the standard star before or after the observations of the target star 
in the same manner. The observation parameters are summarized in Table~\ref{tbl1}.

For the data reduction, we used our own reduction tools and IRAF. 
The standard chop-nod pair subtraction and the shift-and-add method 
in the unit of 0.1 pixel were employed. 
We applied air mass correction by estimating the difference 
in atmospheric absorption using the ATRAN software \citep{lord92}.
The difference in the air mass between the object and the standard star 
was quite small ($\lesssim 0.1$) 
and thus the correction factor in each band is less than $5$~\%.

%% In this section, we use  the \subsection command to set off
%% a subsection.  \footnote is used to insert a footnote to the text.


%% This section also includes several of the displayed math environments
%% mentioned in the Author Guide.

\section{Results}

\subsection{Spectral Energy Distribution}

\begin{figure}
\epsscale{1.0}
\plotone{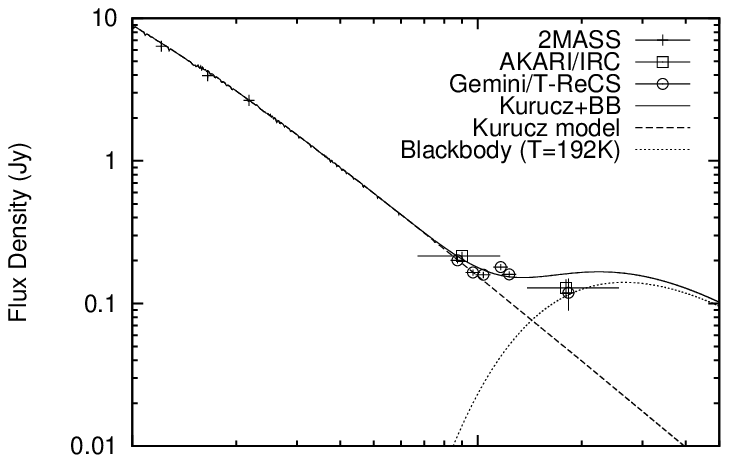}
\plotone{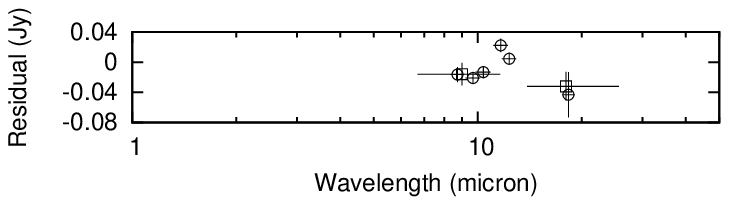}
\caption{{\it Top}: The NIR and MIR SED of HD~106797 
and the results of SED fitting with a model which is consistent with 
a Kurucz model for photospheric contribution and a blackbody of a single temperature 
$T_{\rm d}=192$~K for dust emission.
Open squares and circles indicate the photometry with the {\it AKARI}/IRC and Gemini/T-ReCS, repectively. 
Solid, dashed and dotted lines indicate the total SED, photospheric contribution, 
and dust emission of the blackbody model, respectively. 
{\it Bottom}: The residuals subtracted by the best-fit SED model in the MIR region. 
\label{fig1}}
\end{figure}

The observed flux densities of HD~106797 in all bands are shown in Table~\ref{tbl2}. 
The photospheric flux densities are estimated from the Kurucz model of A0 stars
with the effective temperature of $T_{\rm eff}=9750~{\rm K}$ and 
the surface gravity of $\log g=+4.0$ \citep{kurucz92} 
fitted to the 2MASS $K_{\rm s}$-band photometry of the star and also shown in Table~\ref{tbl2}.
The SED of the star 
in the near-infrared (NIR) and MIR regions is shown in Figure~\ref{fig1}.

No significant excess emission in the {\it AKARI}/IRC S9W band is found. 
The $18~\micron$ flux densities derived with {\it AKARI}/IRC and Gemini/T-ReCS 
are in agreement with each other. 
Although {\it AKARI} data with a beam size of $\sim 5~\arcsec$ 
might be contaminated by other nearby sources, 
we did not find any other infrared sources besides HD~106797 
in the field of view of T-ReCS ($28\farcs8 \times 21\farcs6$).
Therefore it can securely be concluded that 
the $18~\micron$ excess towards HD~106797 is associated with the star.

In addition to the $18~\micron$ excess, T-ReCS narrow band photometry also indicates 
excess emission at 11.7 and $12.3~\micron$ towards HD~106797, 
suggesting that there are hot ($T_{\rm d} \gtrsim 200$~K) debris dust grains around HD~106797. 
% Only about 10 stars with $18~\micron$ excesses are known so far. 
% HD~106797 is a new example and this discovery is very important to study inner debris dust. 
We can also see a bump around $11.7~\micron$ in the SED, 
suggesting the presence of a silicate dust feature in the excess emission.  

To make an initial estimation of the dust temperature and the luminosity, 
we performed a fit with the SED model of 
\begin{eqnarray}
F_{\nu, {\rm model}}(\lambda) = {\rm Kurucz}(T_{\rm eff}=9750~{\rm K}) + {\rm BB}_\nu(\lambda, T_{\rm d}), 
\end{eqnarray}
where ${\rm Kurucz}(T_{\rm eff}=9750~{\rm K})$ is the Kurucz model of A0 stars \citep{kurucz92} 
for the photospheric contribution and ${\rm BB}_\nu(\lambda, T_{\rm d})$ 
is a blackbody of a single temperature $T_{\rm d}$ for excess emission. 
The Kurucz template is scaled to fit the 2MASS $K_{\rm s}$-band flux 
because interstellar extinction is smaller in the $K_{\rm s}$-band 
than in the $J$- and $H$-bands.
The stellar luminosity is derived as $42.4~L_\odot$.
A blackbody of $T_{\rm d}=192$~K gives the best fit to the observed SED 
in the {\it N}- and {\it Q}-bands.
The dust luminosity is derived as $0.00819~L_\odot$.
The dust temperature is appropriate for dust grains at a distance 
of 13.7~AU from the central A0~V star 
when the dust grains are assumed as blackbody that follows 
the relation of $d \propto T_{\rm d}^{-2}$. 
In other words, the inner radius of the debris disk around HD~106797 is $\sim 14$~AU. 
We cannot estimate the degree of extension of the debris disk only from the SED 
since far-infrared (FIR) photometric data are not available. 

Here we performed a fit to the observed excess emission 
with a simple SED model of a blackbody of a single temperature. 
We should note that the flux densities of the best-fit SED model 
at $\lambda \lesssim 10.4~\micron$ are larger than the T-ReCS observations. 
However this problem may be solved by considering dust species whose emissivity 
is small at $\lambda \lesssim 10.4~\micron$ as a carrier of the excess emission. 
We discuss the possible carrier of the excess emission in Section~\ref{Features}.

\subsection{Radial Profile}

We compare the peak-normalized azimuthally averaged radial profile of HD~106797 
at $11.7~\micron$ with the PSF standard by using the T-ReCS data 
to investigate the spatial distribution of the debris dust around HD~106797. 
The debris disk is unresolved and no significant structures 
are seen in the radial profiles in all bands. 
Since the FWHM of the PSF standard at $11.7~\micron$ is $0\farcs43$, 
the size of the debris disk around HD~106797 is less than 
41~AU at the distance of the star ($d=96$~pc),  
which is consistent with the inner radius of the disk estimated from the SED.

\section{Discussion}

\subsection{Features in the {\it N}-band} \label{Features}

\begin{figure}
\epsscale{1.0}
\plotone{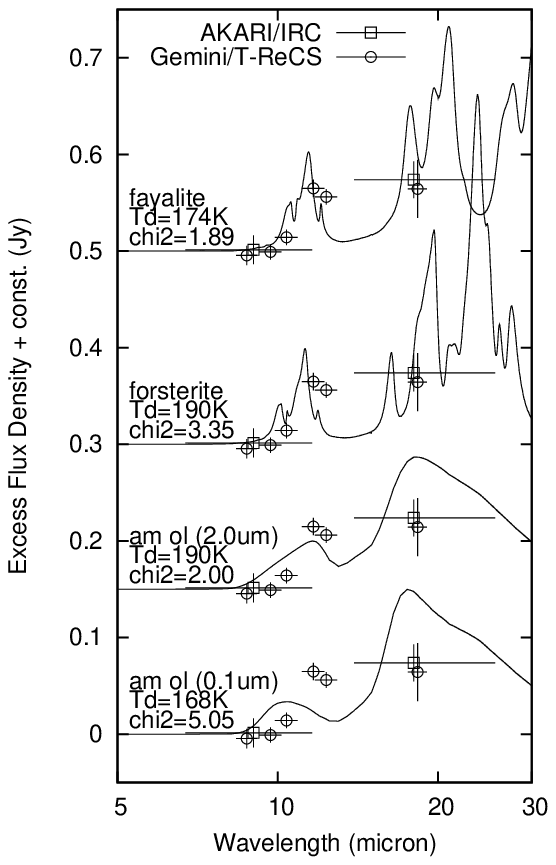}
\caption{The excess flux densities together with model fit results 
of various dust species.
Open squares and circles indicate the residual flux densities 
with the {\it AKARI}/IRC and Gemini/T-ReCS, repectively. 
The solid line indicate the best-fit result for the each dust species 
($0.1~\micron$- and $2.0~\micron$-sized amorphous olivine \citep{dorschner95}, 
crystalline forsterite \citep{koike03}, and crystalline fayalite \citep{koike03}). 
The resultant best-fit dust temperature and $\chi^2_\nu$-value for each dust species are also shown. 
As seen in this figure, crystalline fayalite provides the best-fit result.
\label{fig2}}
\end{figure}

The excess flux densities derived by subtraction of the photospheric contribution 
are shown in Figure~\ref{fig2}. 
The IRC/S9W excess flux density is in agreement with the T-ReCS data 
at $\lambda \lesssim 10.4~\micron$. 
However, the excess flux densities at 11.7 and $12.3~\micron$ are significantly larger 
than those at $\lambda \lesssim 10.4~\micron$, 
suggesting the dust emission has a feature around $11-12~\micron$. 
Since the IRC/S9W band does not cover wavelengths $\gtrsim 11.6~\micron$, 
the 11.7 and $12.3~\micron$ data are not incompatible with the IRC/S9W flux. 

To identify the possible carrier of the feature, 
we perform additional fits for the excess emission with the SED model of 
\begin{eqnarray}
F_{{\rm excess}, \nu}(\lambda) = a \kappa(\lambda) {\rm BB}_\nu(\lambda, T_{\rm d}), 
\end{eqnarray}
where $\kappa(\lambda)$ is mass absorption coefficient of dust and $a$ is scaling factor. 
We consider the mass absorption coefficients of four kinds of silicates, 
$0.1~\micron$- and $2.0~\micron$-sized amorphous olivine \citep{dorschner95}, 
crystalline forsterite, and crystalline fayalite \citep{koike03}. 
The best-fit result for each dust species is overlaid on the observed data 
in Figure~\ref{fig2}. 
Amorphous olivine particles of $0.1~\micron$ size show a triangular feature 
with a peak at $9.7~\micron$ and cannot account for the observed narrow feature 
around $11-12~\micron$. 
Amorphous olivine particles of $2.0~\micron$ size give a better fit than those of $0.1~\micron$ size. 
However they produce extra emission at $\lambda \lesssim 10.4~\micron$, 
and thus they also cannot account for the observations very well.
In contrast, crystalline forsterite and fayalite show the strongest feature at $11.3-11.4~\micron$ 
in the {\it N}-band \citep{koike03}, which may account for the observed narrow feature. 
Polycyclic aromatic hydrocarbons (PAHs) also show a significant feature at $11.3~\micron$ \citep{allamandola89}. 
However PAHs commonly show a stronger feature at $7.7~\micron$ than that at $11.3~\micron$. 
We cannot see any significant excess around $8~\micron$ toward HD~106797, 
and thus PAHs are ruled out as a carrier of $11-12~\micron$ feature. 
Crystalline forsterite and fayalite are likelier carriers of the $11-12~\micron$ feature 
than amorphous silicate and PAHs. 
The resultant $\chi^2_\nu$-values by the fits suggest that fayalite is more plausible. 
The dust temperatures derived from the fits with crystalline forsterite and fayalite 
are 190 and 174~K, respectively. 
Therefore, we conclude that the presence of hot debris dust around the star is secure. 
The $11.3-11.4~\micron$ fine structure is discovered towards some debris disks 
in MIR spectra obtained with the Infrared Spectrograph onboard {\it Spitzer} \citep{beichman05,chen06}
and the COMICS onboard the Subaru Telescope \citep{honda04}, 
which is attributed to crystalline forsterite or fayalite. 
We note that the observed {\it N}-band feature towards 
HD~106797 is not accounted for completely by crystalline forsterite nor fayalite. 
The observed feature seems to be located at slightly longer wavelengths 
than the $11.3-11.4~\micron$ peak of the crystalline forsterite or fayalite, 
suggesting a possible existence of other species of dust grains around HD~106797.

\subsection{Origin of hot dust}

The fractional luminosity $L_{\rm dust}/L_{\rm star}$ of HD~106797 
is derived as $\sim 1.93 \times 10^{-4}$ 
by integrating the best-fit SED model of the star and the excess from 0.01 to $1000~\micron$. 
The fractional luminosity of our own zodiacal cloud is estimated 
as an order of $\sim 10^{-7}$ \citep{backman93}. 
Thus the debris disk around HD~106797 is 1000 times brighter than our own zodiacal cloud. 
\citet{wyatt07} develop a simple model for the steady state evolution of debris disks 
due to collisions based on \citet{wyatt02}. 
The model indicates that a fractional luminosity larger than $10^{-4}$ is 
obtained only in the stage less than a few~Myr for A0~V stars.
Although the age of HD~106797 is unknown, the star is thought to be older than 10~Myr 
since the star is not labeled as an emission line star in 
the Tycho-2 spectral type catalog \citep{wright03}, 
which is one of the strong indicators of young stars. 
Therefore the existence of a large amount of hot dust around HD~106797 cannot be 
accounted for by a simple steady state model. 
The system must be undergoing some kind of transient events. 
For example, the origin for this transient dust may be accounted for by a dynamical instability 
%% caused by migration of gas giant planet 
that scatters planetesimals inward from a more distant planetesimal belt.  
Then the dust is released from the planetesimals in collisions and sublimation. 
The transient event is akin to the late heavy bombardment (LHB) in the solar system, 
a cataclysmic event 700 Myr after the initial formation of the solar system, 
as implied by the Moon's cratering record \citep[e.g.][]{hartmann00}. 

As discussed in a previous section, 
HD~106797 shows a $11-12~\micron$ feature that may originate from crystalline silicate. 
In the interstellar medium, silicates are mostly amorphous \citep{kemper04}.
Crystallization of silicates requires heating to a temperature higher than 800~K \citep{hallenbeck00}. 
How were high temperature products such as crystalline silicate produced in the region 
at which dust temperature is estimated to be less than 300~K? 
This question is akin to the crystalline silicate problem of comets in the solar system. 
Several models have been proposed to account for the problem. 
\citet{bockelee02} propose a model in which silicate dust particles heated by radiation 
from the central star and crystallized at the central region of the disk are transported 
outward to the cold region of the disk by a turbulent flow or an X-wind. 
On the other hand, \citet{harker02} propose a model in which silicate dust particles 
in the cold region of the outer disk are heated by shock wave and crystallized in situ. 
Both models are able to account for the distributions of crystalline silicate 
in a few tens of AU from the central star 
although it is still unclear which mechanism is the most effective. 
Resolving the radial distributions of every dust species in the debris disk of HD~106797
with future spectroscopy with high spatial resolution 
should give us an insight into the problem.

While more than one hundred debris disks 
with large $60~\micron$ excess have been discovered \citep{rhee07}, 
only seven debris disks with large $10~\micron$ excess 
(fractional luminosity $\gtrsim 0.5$ at $10~\micron$) have been reported so far
\citep[$\beta$~Pic, HIP~8920, HD~113766, HR~7012, $\eta$~Crv, HD~145263, and HD~202406;][]{telesco05,song05,chen06,smith08}. 
Therefore debris disks with large $10~\micron$ excess are so far rare.
All of the $10~\micron$ excesses are thought to attribute 
to hot dust of $T_{\rm d} \gtrsim 200$~K.
$\beta$~Pic also shows large far-infrared excess 
attributable to abundant cold dust \citep{backman92}. 

In should be noted that in spite of the large MIR excess, 
HD~106797 is not detected in the FIR region either by the Far-Infrared Surveyor (FIS)  
onboard {\it AKARI} (J.~P.~Marshall, private communication) or by the {\it IRAS} observations. 
Both of the the FIS/WIDE-S ($60-110~\micron$) 
and the {\it IRAS} $60~\micron$ detection limits are about $1.5$~Jy. 
Thus HD~106797 may be an example of a debris disk source,  
in which hot dust is very abundant while Kuiper-belt-analog cold dust is not. 
The present discovery of the $10~\micron$ excess towards HD~106797 
suggests a presence of the new kind of debris disk around main-sequence stars. 

Most of the reported debris disks with large $10~\micron$ excess including HD~106796 
show fine structures in the {\it N}-band attributable to crystalline silicates 
\citep{knacke93,song05,chen06,honda04} 
except for HD~202406, whose MIR measurements with a wavelength resolution 
high enough to discuss the fine structure are not available. 
We speculate that transient events like the LHB tend to cause dust heating 
and generate crystalline silicates efficiently.

\acknowledgments

This research is based on observations with the {\it AKARI}, 
a JAXA project with the participation of ESA. 
This research is also based on data collected at the Gemini Observatory, 
through the time exchange programs with the Subaru Telescope, 
which is operated by the National Astronomical Observatory of Japan.  
We appreciate the support from the Gemini Observatory staff. 
We thank Chiyoe Koike and Hiroki Chihara 
for providing us with crystalline silicate spectra and their useful comments. 
We also thank the anonymous referee, Aki Takigawa, Shogo Tachibana, 
and Alexander V. Krivov for their useful comments and suggestions.
This research was supported by the MEXT, ``Development of Extrasolar Planetary Science,'' 
and the UK science and Technology Facilities Council. 
H.F. is financially supported by the Japan Society for the Promotion of Science.

{\it Facilities:} \facility{{\it AKARI} (ISAS/JAXA)}, \facility{Gemini-South (AURA)}.

\clearpage

%% Table 1

\begin{table}
\begin{center}
\caption{Summary of Gemini/T-ReCS observations.\label{tbl1}}
\begin{tabular}{cccccc}
\tableline\tableline
Object & Filter      & Date \& Time & Integration & Air Mass & Comment \\
       & ($\micron$) & (UT)         & (s)         &          &         \\
\tableline
HD~106797 & 18.3     & 10 June 2008 04:18:04 & 811 & 1.79    & \nodata  \\
HD~110458 & 18.3     & 10 June 2008 05:09:19 & 116 & 1.91    & Standard \\
HD~106797 & 12.3     & 12 June 2008 03:10:34 & 174 & 1.50    & \nodata  \\
HD~106797 & 11.7     & 12 June 2008 03:20:49 & 116 & 1.52    & \nodata  \\
HD~106797 & 10.4     & 12 June 2008 03:27:48 &  58 & 1.54    & \nodata  \\
HD~106797 & 9.7      & 12 June 2008 03:31:32 & 174 & 1.56    & \nodata  \\
HD~106797 & 8.8      & 12 June 2008 03:41:48 &  58 & 1.59    & \nodata  \\
HD~110458 & 8.8      & 12 June 2008 03:48:42 &  58 & 1.45    & Standard \\
HD~110458 & 9.7      & 12 June 2008 03:52:26 &  58 & 1.46    & Standard \\
HD~110458 & 10.4     & 12 June 2008 03:56:10 &  58 & 1.48    & Standard \\
HD~110458 & 11.7     & 12 June 2008 03:59:55 &  58 & 1.50    & Standard \\
HD~110458 & 12.3     & 12 June 2008 04:03:39 &  58 & 1.52    & Standard \\
\tableline
\end{tabular}
%\tablecomments{We can also attach a long-ish paragraph of explanatory material to a table.}
\end{center}
\end{table}

% \clearpage

%% Table 2

\begin{table}
\begin{center}
\caption{Infrared Photometry of HD~106797.\label{tbl2}}
\begin{tabular}{ccccc}
\tableline\tableline
$\lambda$   & $F_\nu$      & Instrument & Photosphere\tablenotemark{a} & Significance, $\chi$\tablenotemark{b} \\
($\micron$) & (mJy)        &            & (mJy)   \\
\tableline
8.8         & $201 \pm 20 $ & T-ReCS  & 206 & -0.3 \\
9           & $216 \pm 15$  & IRC     & 214 &  0.1 \\
9.7         & $165 \pm 17 $ & T-ReCS  & 166 & -0.1 \\
10.4        & $159 \pm 15 $ & T-ReCS  & 145 &  0.9 \\
11.7        & $180 \pm 14 $ & T-ReCS  & 115 &  4.6 \\
12.4        & $160 \pm 13 $ & T-ReCS  & 104 &  4.2 \\
18          & $128 \pm 19$  & IRC     & 55  &  3.8 \\
18.3        & $119 \pm 59 $ & T-ReCS  & 47  &  1.2 \\
\tableline
\end{tabular}
\tablenotetext{a}{From Kurucz model to fitted to 2MASS Ks-band data.}
\tablenotetext{b}{$\chi = ({\rm Observed} - {\rm Kurucz}) / {\rm noise}$.}
%\tablecomments{We can also attach a long-ish paragraph of explanatory material to a table.}
\end{center}
\end{table}

\clearpage

%% Figure 1

% \clearpage

%% Figure 2

\clearpage

\end{document}